# Spin/Valley pumping of resident electrons in WSe$_2$ and WS$_2$ monolayers

Cedric Robert[1], Sangjun Park[2], Fabian Cadiz[2], Laurent Lombez[1], Lei Ren[1], Hans Tornatzky[1], Alistair Rowe[2], Daniel Paget[2], Fausto Sirotti[2], Min Yang[3], Dinh Van Tuan[3], Takashi Taniguchi[5], Bernhard Urbaszek[1], Kenji Watanabe[5], Thierry Amand[1], Hanan Dery[3,4] and Xavier Marie[1]

[1]Université de Toulouse, INSA-CNRS-UPS, LPCNO, 135 Av. Rangueil, 31077 Toulouse, France
[2]Physique de la matière condensée, Ecole Polytechnique, CNRS, IP Paris, 91128 Palaiseau, France
[3]Department of Electrical and Computer Engineering, University of Rochester, Rochester, New York 14627, USA
[4]Department of Physics, University of Rochester, Rochester, New York 14627, USA
[5]National Institute for Materials Science, Tsukuba, Ibaraki 305-004, Japan

*Monolayers of transition metal dichalcogenides are ideal materials to control both spin and valley degrees of freedom either electrically or optically. Nevertheless, optical excitation mostly generates excitons species with inherently short lifetime and spin/valley relaxation time. Here we demonstrate a very efficient spin/valley optical pumping of resident electrons in n-doped WSe$_2$ and WS$_2$ monolayers. We observe that, using a continuous wave laser and appropriate doping and excitation densities, negative trion doublet lines exhibit circular polarization of opposite sign and the photoluminescence intensity of the triplet trion is more than four times larger with circular excitation than with linear excitation. We interpret our results as a consequence of a large dynamic polarization of resident electrons using circular light.*

Transition metal dichalcogenides (TMD) such as MoS$_2$, MoSe$_2$, WS$_2$, or WSe$_2$ are layered semiconductors with promising applications in optoelectronics and spintronics [1]. In the monolayer (ML) limit, they become direct band gap semiconductors, with gaps located at the six corners of the hexagonal Brillouin zone (K valleys) [2–4]. Remarkably, they exhibit a strong light-matter interaction governed by tightly bound excitons with binding energies of several hundreds of meV [5]. In addition, they are characterized by a strong spin-orbit coupling and a lack of crystal inversion symmetry resulting in original spin/valley properties [6–9]. Among them, chiral optical selection rules dictate that circularly polarized light can photo-generate carriers in either K or K' valleys with either spin up or spin down, i.e. the so-called spin/valley pumping. Thus, TMD MLs were quickly considered as an ideal platform to control both spin and valley degrees of freedom with potential applications in quantum information processing [10–13]. Nevertheless, light excitation usually yields neutral excitons and using these photo-generated species to encode spin or valley information is inherently limited by both their short recombination time (~ps) [14–16] and their very fast spin/valley relaxation time induced by electron-hole exchange interaction (~ps) [17,18]. Recently other strategies have been proposed using longer lived excitonic species such as dark excitons, dark trions or interlayer excitons in heterostructures [19–21]. Another promising route consists in using resident electrons or holes in doped monolayers. Beyond its obvious advantage for future devices as compared to the manipulation of excitons, the spin/valley relaxation of resident carriers is prevented by spin/valley locking and is not governed by efficient exchange interaction like for excitons. Spin/valley relaxation times as long as 100's ns to several μs for electrons and holes have been measured in WSe$_2$ using time-resolved Kerr experiments and spin/valley noise spectroscopy [22–25]. Nevertheless, very little is known about the polarization mechanism and the maximum degree of polarization one can reach for resident carriers. Back et al. [26] showed that a near complete valley polarization of electrons can be reached in a n-doped MoSe$_2$ ML but it requires an out-of-plane magnetic field of 7 T that is incompatible with the development of future devices.

In this letter, we demonstrate a very efficient spin/valley pumping mechanism which yields very large polarization for resident electrons in n-doped WSe$_2$ and WS$_2$ following a circularly polarized excitation without applying any magnetic field. In contrast to pump-probe experiments, we use continuous-wave (cw) laser excitation that leads to a dynamical building of this very large polarization. We use the degree of circular polarization of the photoluminescence associated with negative trions as probes of the polarization of electrons (both the intervalley triplet trion $X^{T-}$ and the intravalley singlet trion $X^{S-}$ which consist in the binding of a photo-generated electron-hole pair with a resident electron from the opposite (same) valley (see Figure 3a)). In n-WSe$_2$ we measure a very large positive circular polarization 90% for the triplet trion and a negative polarization −40% for the singlet trion. Remarkably, the total intensity of the triplet trion following circular excitation is more than four times larger than the total intensity following linear excitation. Using simple models of trion formation, we demonstrate that all these observations are consistent with a very efficient spin/valley pumping of resident electrons and give an estimate of ~80% for polarization.

**Results**

We fabricate a high quality WSe$_2$ charge tunable device as sketched in Figure 1a. Details of the sample fabrication can be found in the supplemental materials S1. By tuning the voltage bias between a back gate and the ML we can electrostatically dope the ML. We then perform polarization dependent micro-photoluminescence (PL) experiments in the n-doping regime at a temperature of 4 K. The excitation source is the 632.8 nm line of a HeNe laser. Unless otherwise stated, the excitation power is 5 µW focused to a spot size smaller than 1 µm diameter. We also study a naturally n-doped WS$_2$ ML without charge tuning with a cw 570 nm laser (excitation power of 18 µW and temperature of 20 K). More details on the experimental setup can be found in supplemental materials S1. Importantly, we restrict our study to moderate electron densities of a few $10^{11}$ cm$^{-2}$ so that the simple three particles picture (i.e. trions) is equivalent to the many-body picture (i.e. Fermi polarons) [27,28].

We first present in Figure 1b the PL color plot as a function of bias in the charge tunable WSe$_2$ device. We recognize several exciton species including the bright neutral exciton ($X^0$), the dark neutral exciton ($X^D$), the bright trions (intervalley triplet $X^{T-}$ and intravalley singlet $X^{S-}$) and the dark trion ($X^{D-}$), in agreement with previous studies [29–37]. The spectral linewidth of $X^0$ at the neutrality point is as low as 2.5 meV (FWHM) vouching for the state-of-the-art quality of the sample [34]. In Figure 1c, we also present the color plot of the reflectivity contrast highlighting transitions with large oscillator strength (i.e. $X^0$, $X^{T-}$ and $X^{S-}$). In the following we will focus on an electron doping density of $4 \times 10^{11}$ cm$^{-2}$ (see white dashed line in Figure 1b and c) where $X^{T-}$ and $X^{S-}$ dominate the PL spectrum.

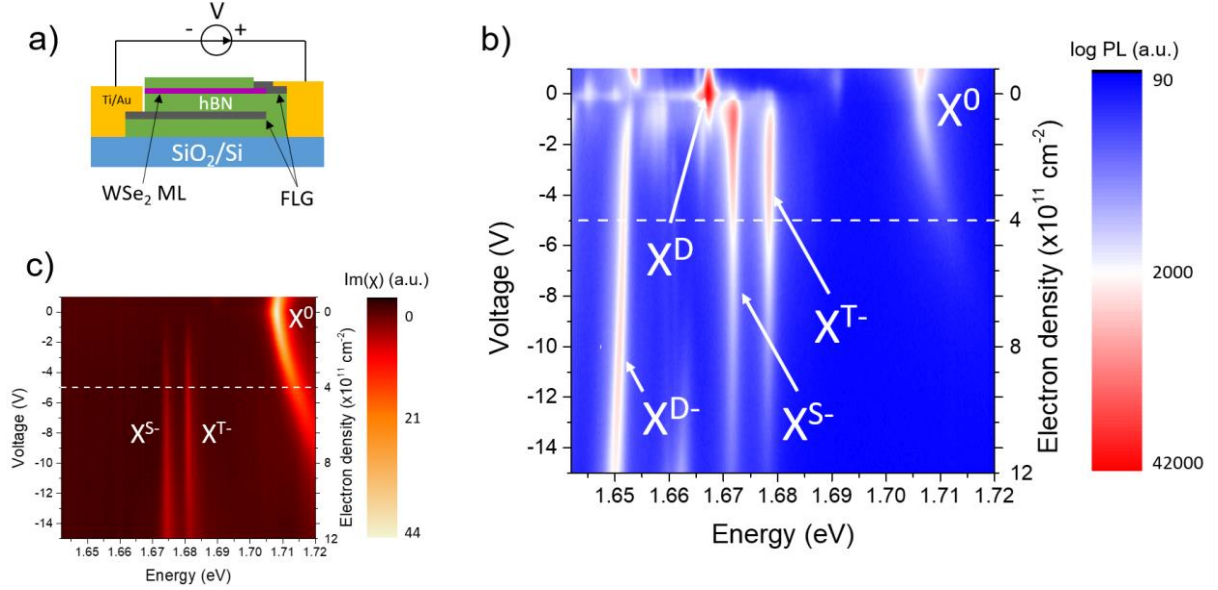

*Figure 1: (a) Sketch of the sample. (b) PL intensity as a function of electron density. The excitation energy is 1.96 eV. (c) Imaginary part of the optical susceptibility χ measured using differential reflectivity as a function of electron density. The reflectivity contrast is transformed into Im(χ) using a Kramers-Kronig transform [26].*

Figure 2 presents the key results of this work. In Figure 2a, we show the photoluminescence spectra for both σ+ and σ- detections following σ+ excitation. We define the degree of circular polarization as $P_c = \frac{I_{\sigma+}-I_{\sigma-}}{I_{\sigma+}+I_{\sigma-}}$ where $I_{\sigma+}, I_{\sigma-}$ are the PL intensities with σ+ and σ- detection respectively. While the bright exciton $X^0$ exhibits a positive circular polarization below +20% as a consequence of the efficient long-range exchange interaction, the lines of the bright trion doublet show strong polarization of opposite sign: +91% for $X^{T-}$ and −40% for $X^{S-}$ at the peak. Note that this negative polarization on the singlet has been observed elsewhere recently in state-of-the-art samples [34]. The dark trion $X^{D-}$ shows no circular polarization in agreement with its out of plane polarization [31,36,37].

Then we switch to linear excitation $\pi_X$ and measure both co-linear $I_X$ and cross-linear $I_Y$ intensities. We define the total PL intensity following linear excitation as $I_\pi = I_X + I_Y$ and the total intensity following circular excitation as $I_\sigma = I_{\sigma+} + I_{\sigma-}$. We show in Figure 2b both $I_\pi$ and $I_\sigma$ and the ratio $R = \frac{I_\sigma}{I_\pi}$. There is no difference in intensity for $X^0$ between linear and circular excitation (i.e. $R$=1). On the other hand, R reaches a very large value of 4.4 at the peak of $X^{T-}$ and slightly below 1 for $X^{S-}$ and $X^{D-}$. In other words and surprisingly, the PL intensity of $X^{T-}$ is more than 4 times larger when we excite with circularly polarized light.

In Figure 2c and d, we show $P_c$ and $R$ measured at the emission peaks of the bright trion doublet as a function of electron density. The very large positive polarization of $X^{T-}$ is nearly constant while for $X^{S-}$ it varies from positive at small doping to negative for densities above $2\times10^{11}$ cm$^{-2}$ and reaches the minimum value of −40% for $4\times10^{11}$ cm$^{-2}$. Concerning the ratio of PL intensities between circular and linear excitations (Figure 2d), it remains above $R$=2 in the whole investigated electron density range for $X^{T-}$ and slightly below 1 for $X^{S-}$. Finally, we present in Figure 2e and f the excitation power dependence at an electron doping density of $3.2\times10^{11}$ cm$^{-2}$. We clearly see that when we reduce the excitation power, $P_c$ converges to a value around 50% for both $X^{T-}$ and $X^{S-}$ and that $R$ decreases and gets closer to 1 for $X^{T-}$, while it stays constant and close to 1 for $X^{S-}$.

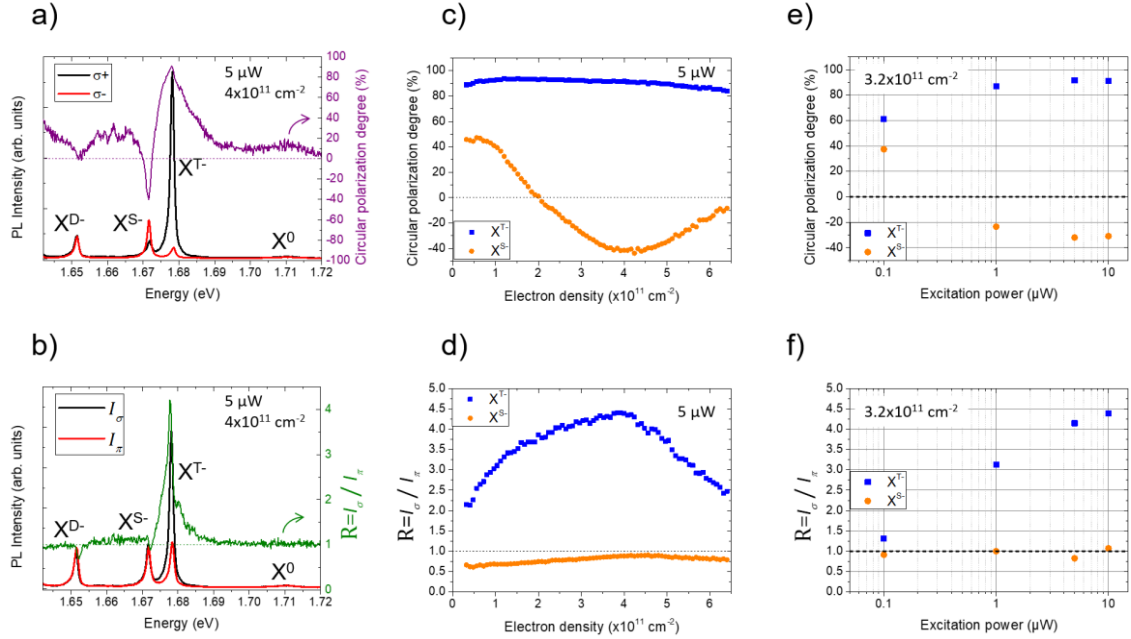

*Figure 2: (a) Photoluminescence and circular polarization spectra for σ+ and σ- detections with σ+ excitation. (b) Total photoluminescence spectra with circular excitation and linear excitation. (c) Circular polarization degree at the peak of triplet and singlet negative trions as a function of electron density. (d) Ratio of PL intensities between circular and linear excitations at the peak for both triplet and singlet as a function of electron density. (e) Circular polarization degree at the peak of triplet and singlet as a function of excitation power. (f) Ratio of PL intensities between circular and linear excitations at the peak for both triplet and singlet as a function of excitation power.*

**Discussion**

In the following, we will tentatively explain these results focusing on three clear observations:

(i) the circular polarization of the triplet trion can reach very high positive values.

(ii) the circular polarization of the triplet and singlet trions are of opposite sign at sufficiently large doping level.

(iii) the intensity of the triplet trion is more than 4 times larger with circular excitation than with linear excitation.

We show in Figure 3 the three-particle configurations of σ+ and σ- triplet and singlet trions. A triplet trion consists of a photo-generated electron-hole pair (exciton made of an electron in the topmost conduction band and a missing electron in the same valley) bound to a resident electron in the bottom conduction band lying in the other valley. On the other hand, a singlet trion is composed of a photo-generated electron-hole pair bound to a resident electron in the same valley. Experimentally, we observe that when exciting with a σ+ polarized laser the two strongest PL peaks are the σ+ triplet trion and the σ- singlet trion (Figure 2a). These two configurations are highlighted in red in Figure 3a. In both cases the resident electron in the three particle complex lies in the K' valley. Thus if we assume that the formation mechanisms of triplet and singlet trions are the same, the opposite sign of polarization of triplet and singlet trions can only be explained by a larger population of resident electrons in the K' valley as compared to the K valley; i.e. by spin-valley pumping of resident electrons with spin up in K' valley using σ+ polarized light.

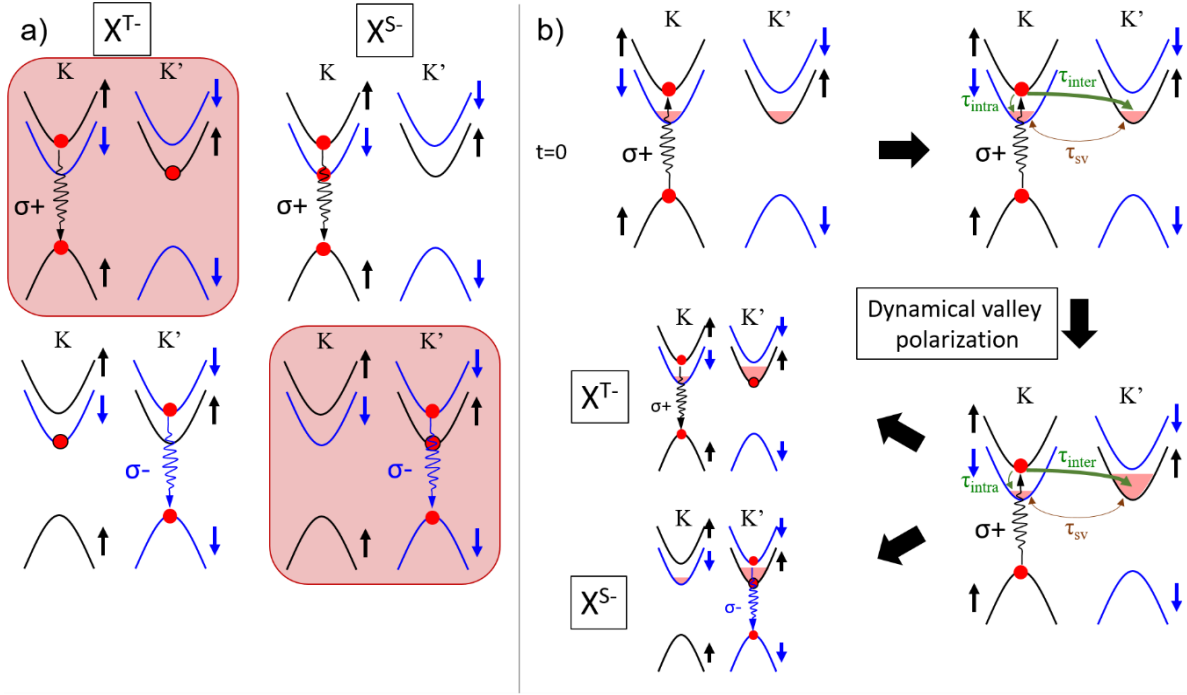

*Figure 3: (a) Sketches of the three particle pictures of bright triplet $X^{T-}$ and singlet $X^{S-}$ negative trions with σ+ and σ- emission. The two majority trions following a σ+ cw-excitation are highlighted in red. (b) Sketch of the dynamic polarization of resident electrons with σ+ excitation. $\tau_{inter}$ ($\tau_{intra}$) represents the intervalley spin-conserving (intravalley spin-flip) relaxation times of topmost electrons while $\tau_{sv}$ corresponds to the spin/valley relaxation time of resident electrons.*

We propose in Figure 3b a mechanism that dynamically polarizes the resident electrons, in a manner analogous to dynamic spin polarization in conventional semiconductors [38–40]. Without light excitation and magnetic field, the densities of resident electrons are similar in the K and K' valleys. By continuously exciting with σ+ polarized light at t >0, electrons are photo-generated in the conduction-band top valley of K. These electrons are either free or bound to photo-generated holes and can relax to the bottom conduction bands both in K and K' valleys through different mechanisms. Intravalley relaxation (time constant $\tau_{intra}$ in Figure 3b) requires an electron spin-flip whereas intervalley relaxation (time constant $\tau_{inter}$ in Figure 3b) conserves the electron's spin. When energy relaxation is governed by electron-phonon interactions, spin-conserving relaxation is associated to the gradient of the spin-independent component of the crystal potential, while spin-flip interactions are associated to the spin-orbit interaction component. As long as the electronic states are not strongly spin-mixed, the spin conserving processes are typically stronger, resulting in faster momentum relaxation compared with spin relaxation (i.e. $\tau_{inter} < \tau_{intra}$). Recently, He *et al.* analyzed the dark trions' polarization in ML-WSe$_2$ and showed that spin-conserving intervalley relaxation, mediated by zone-edge phonons, is indeed stronger than spin-flip intravalley relaxation that is mediated by zone-center phonons [34]. Consequently, the electron population in the bottommost K' conduction band becomes larger than the electron population in the bottommost K conduction band upon excitation by a circularly polarized light σ+ (i.e., valley polarization). Under cw-excitation, the mechanism of Figure 3b is repeated multiple times resulting in a dynamical buildup of valley polarization. This dynamical valley polarization is sustainable if the generation rate of photo-excited electron-hole pairs is faster than the rate at which electrons reestablish thermal equilibrium between the bottommost conduction-band valleys of K and K' through intervalley spin-relaxation (time constant $\tau_{sv}$ in Figure 3b). The latter is a relatively slow process at low temperatures, measured to be as long as 100's ns to several µs in ML-WSe$_2$ [22–25], because it is mediated by spin-flip intervalley transitions, which are forbidden to leading order by time-reversal symmetry [41,42]. Thus, even if the buildup of dynamical valley polarization is slow because the intervalley spin-conserving relaxation ($\tau_{inter}$) is not much faster than the intravalley spin-flip one ($\tau_{intra}$), the attainable valley polarization can still be very large (we will give an estimate of ~−80% in the following). We mention that other mechanisms of polarization transfer from photo-generated carriers to resident electrons have been proposed by Ersfeld et al. [43] and Fu et al. [44] considering differences in the recombination rates of indirect excitons and spin-forbidden dark excitons or differences in the relaxation rates of singlet and triplet to the dark trions. In each scenario, the two ingredients are the same: creation of an asymmetry in the population of resident electrons and long spin-flip intervalley relaxation times.

Once we consider that resident electrons mainly populate the K' valley, we can explain the very large polarization of the triplet trion and the negative polarization of the singlet trion. We first assume that the bright trions are formed through the binding of photo-generated bright excitons with a resident electron (i.e. a bimolecular formation [45]). We will discuss other possible mechanisms in a next section and in the Supplemental Materials. We also assume that the electron density is much larger than the photo-generated exciton density (see supplemental materials S2 for our estimation of the exciton density) and we assume that the spin relaxation of trions is much slower than their recombination lifetimes (i.e. the observed polarization in cw experiments correspond to the polarization at the trion formation, see supplemental materials S3 for the justification of this assumption). In this case we can calculate the polarization of triplet and singlet trions as a function of the polarization of resident electrons $P_e = \frac{n_e^K - n_e^{K'}}{n_e^K + n_e^{K'}}$ (where $n_e^K$ and $n_e^{K'}$ are the populations of resident electrons in the K and K' valleys) and the polarization of photo-generated excitons $P_0 = \frac{N_0^K - N_0^{K'}}{N_0^K + N_0^{K'}}$ (where $N_0^K$ and $N_0^{K'}$ are the populations of photo-generated excitons in the K and K' valleys) (see supplemental materials S4 for more details):

$$\underline{\text{Triplet}} \qquad P_c(X^{T-}) = \frac{P_0 - P_e}{1 - P_0 P_e} \qquad (1)$$

$$\underline{\text{Singlet}} \qquad P_c(X^{S-}) = \frac{P_0 + P_e}{1 + P_0 P_e} \qquad (2)$$

The results of Figure 2a ($P_c(X^{T-})$=91% and $P_c(X^{S-})$=−40%) match well with $P_0$=51% and $P_e$=−76%; i.e. the resident electrons are strongly polarized in the K' valley.

This simple scenario of dynamic polarization of electrons is consistent with the power dependence of Figure 2e. Indeed, when the excitation power decreases, the polarization of both $X^{T-}$ and $X^{S-}$ converge to the same value of around +50%. In this case, the photo-generation rate of electrons is not sufficient to create a significant polarization of resident electrons. Thus the polarizations of $X^{T-}$ and $X^{S-}$ mainly reflect the polarization of the exciton reservoir just before the formation of trions (i.e. the polarization of the hot excitons $P_0$). Furthermore, the doping density dependence of trions circular polarization of Figure 2c can be qualitatively explained. For doping densities above $4 \times 10^{11}$ cm$^{-2}$, the polarization in absolute value of both $X^{T-}$ and $X^{S-}$ drops because the density of photo-generated electrons is not large enough to fully polarize the resident electrons. In supplemental materials S5, we show that increasing the excitation power results in larger polarizations for larger doping densities.

We now discuss the third main result of this work which is another consequence of the efficient spin-valley pumping of resident electrons: the PL intensity of the triplet trion is stronger with circular excitation than with linear excitation (Figure 2b). Note that this characteristic has been observed in GaAs-based alloys (GaAsN, GaAlAs) where it was attributed to spin dependent recombination via paramagnetic centers [46–48]. Here we attribute it to the efficient spin-valley pumping of resident electrons. Considering the same simple model based on the bimolecular formation of trions that we used to calculate the degrees of circular polarization, we can show that the ratio of PL intensities between circular and linear excitation are (details of the calculations are presented in the supplemental materials S4):

<u>Triplet</u> $\qquad\qquad\qquad R(X^{T-}) = 1 - P_0 P_e \qquad\qquad (3)$
<u>Singlet</u> $\qquad\qquad\qquad R(X^{S-}) = 1 + P_0 P_e \qquad\qquad (4)$

Using the values $P_0=51\%$ and $P_e=-76\%$ as determined previously we get qualitative agreement with our experimental results: the PL intensity of the triplet trion is larger with circular excitation (i.e. $R(X^{T-})=1.39 > 1$) and the PL intensity of the singlet trion is larger with linear excitation $R(X^{S-})=0.61 < 1$.

Nevertheless, our simple model does not describe three quantitative aspects:

- (i) the ratio of the total intensities (triplet + singlet: $R(X^{T-} + X^{S-})$ using this model is equal to 1 while it is clearly larger than 1 in Figure 2b.

- (ii) $R(X^{T-})$ cannot be larger than 2 in equation (3) while it is experimentally larger than 4.

- (iii) The circular polarization of the singlet trion turns positive at low doping in Figure 2c.

The first limitation suggests that considering the subspace triplet+singlet is insufficient to fully explain our results (i.e. we have a deficit of luminescence for linear excitation). As shown in Figure 2b, we clearly see that the ratio $R(X^{D-})$ for the dark trion transition is also below 1; i.e. more intensity with linear excitation than with circular excitation. The dark trion formation path should thus be included.

The two other limitations (ii) and (iii) suggest alternative mechanisms for the formation of the trion species (bright and dark). Theoretical and experimental studies on the trion formation processes in TMD MLs are very scarce. Singh *et al.* measured the bright trion formation time in MoSe$_2$ ML using resonant excitation at the bright exciton transition energy [49] and linked it to the exciton-electron interaction. Here we use non-resonant excitation above the free carrier band gap of WSe$_2$. We can thus propose different formation mechanisms. The bright trions can be formed through the binding of bright excitons with resident electrons (i.e. the bimolecular process already considered above) but also through the binding of two electrons and a hole (i.e. a trimolecular process). In addition, singlet and triplet trions can be formed through the binding of a topmost conduction band electron with respectively a spin-forbidden dark exciton and a momentum-indirect exciton. Similar mechanisms should also be considered for the formation of dark trions in addition to the possible relaxation from bright trions. The dominant formation processes certainly depends on the doping density. For instance the trimolecular process has been demonstrated as dominant in GaAs quantum wells at sufficiently high doping while the bimolecular one is dominant at lower doping densities [45]. The determination of the trion formation processes is beyond the scope of this paper and will require additional theoretical work. In the supplemental materials S8, we tentatively present a scenario based on trimolecular formation of bright and dark trions that matches quantitatively with the measured values of $P_c(X^{T-})$, $P_c(X^{S-})$, $R(X^{T-})$ and $R(X^{S-})$ at a doping density of $4\times10^{11}$ cm$^{-2}$ and propose some scenarios to explain the positive circular polarization of the singlet trion at low doping.

Finally, we show that the manifestations of efficient spin valley-pumping of resident electrons in WSe$_2$ are also observed in WS$_2$. In Figure 4, we show the PL spectra, the circular polarization degree and the ratio $R = \frac{I_\sigma}{I_\pi}$ for a hBN-encapsulated WS$_2$ ML. In this case, the ML is not gated but it is intrinsically slightly n-doped as proved by the presence of both triplet and singlet negative trions (X$^{T-}$ and X$^{S-}$) in the luminescence spectra [50–52]. The results are very similar to slightly n-doped WSe$_2$: X$^{T-}$ is strongly positively polarized (76% at the peak) and more intense with circular excitation than with linear excitation $R(X^{T-})$=1.85. We do not observe the negative polarization for the singlet as in Figure 2a but $P_c(X^{S-})$ is slightly positive as observed for WSe$_2$ at smaller doping (Figure 2c). Similar power dependence is also observed and presented in the supplemental materials S9.

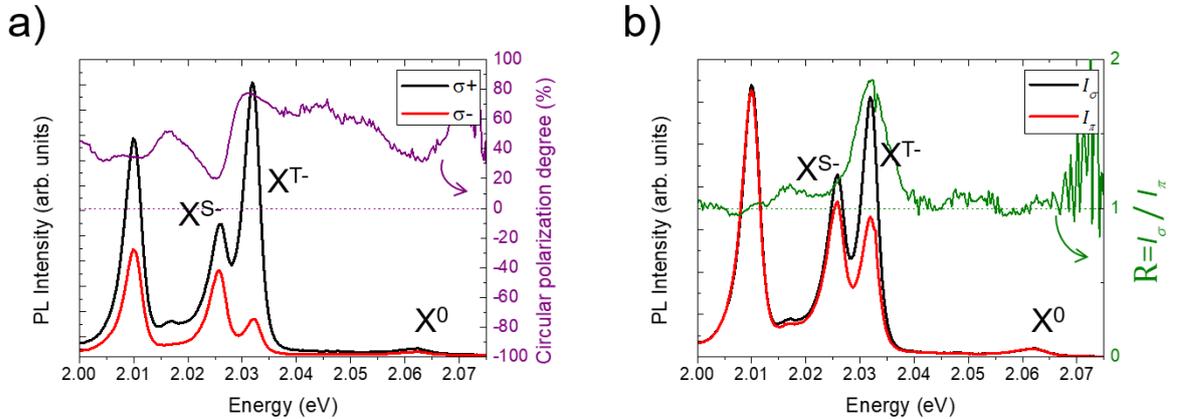

*Figure 4: (a) Photoluminescence spectra of WS$_2$ for σ+ and σ- detection with σ+ excitation. (b) Photoluminescence spectra of WS$_2$ with circular excitation and linear excitation. Excitation power is 18 μW.*

In summary, we demonstrated very efficient spin-valley pumping of resident electrons in both WSe$_2$ and WS$_2$ monolayers using circularly polarized light. This process manifests as a large positive circular polarization of the triplet trion, a negative polarization of the singlet trion and a large increase of the triplet trion PL intensity with circular excitation as compared to linear excitation. Interestingly, these results demonstrate that circularly polarized excitation photo-generates electron-hole pairs in one valley and dynamically polarize resident electrons in the opposite valley. This work is thus an important step towards the development of valleytronic devices based on TMD MLs.

Acknowledgments: We thank Scott A. Crooker, Jing Li and Mateusz Goryca for fruitful discussions. The work at Rochester was funded by the Department of Energy, Basic Energy Sciences, under Contract No. DE-SC0014349. This work was supported by Agence Nationale de la Recherche funding ANR VallEx, ANR 2D-vdW-Spin, ANR MagicValley and ANR SpinCAT (No. ANR-18-CE24-0011-01). K.W. and T.T. acknowledge support from the Elemental Strategy Initiative conducted by the MEXT,Japan and and the CREST (JPMJCR15F3), JST. X.M. also acknowledges the Institut Universitaire de France.

# Supplemental Material for "Spin/Valley pumping of resident electrons in WSe$_2$ and WS$_2$ monolayers"

Cedric Robert[1], Sangjun Park[2], Fabian Cadiz[2], Laurent Lombez[1], Lei Ren[1], Hans Tornatzky[1], Alistair Rowe[2], Daniel Paget[2], Fausto Sirotti[2], Min Yang[3], Dinh Van Tuan[3], Takashi Taniguchi[5], Bernhard Urbaszek[2], Kenji Watanabe[5], Thierry Amand[1], Hanan Dery[3,4] and Xavier Marie[1]

[1]Université de Toulouse, INSA-CNRS-UPS, LPCNO, 135 Av. Rangueil, 31077 Toulouse, France
[2]Physique de la matière condensée, Ecole Polytechnique, CNRS, IP Paris, 91128 Palaiseau, France
[3]Department of Electrical and Computer Engineering, University of Rochester, Rochester, New York 14627, USA
[4]Department of Physics, University of Rochester, Rochester, New York 14627, USA
[5]National Institute for Materials Science, Tsukuba, Ibaraki 305-004, Japan

## S1. Sample fabrication and experimental set-up

We have fabricated a van der Waals heterostructure (sketched in Fig. 1a of the main text) made of an exfoliated ML-WSe$_2$ embedded in high quality hBN crystals [1] using a dry stamping technique [2] in the inert atmosphere of a glove box. The layers are deterministically transferred on top of a SiO$_2$/Si substrate with Ti/Au electrodes patterned by photolithography. Flux-grown WSe$_2$ bulk crystals are purchased from 2D semiconductors. We use few layers of graphene exfoliated from a HOPG bulk crystal for the back gate and to contact the ML-WSe$_2$. The WS$_2$ sample is fabricated using the same technique but without electrodes.

The carrier density in the charge tunable device is estimated by two methods as in our previous work [3]. The first one uses the simple plate capacitance model. Knowing the applied voltage (V), the hBN thickness $t$ (210 nm in our device) and using a hBN dielectric constant of $\varepsilon_{hBN} \sim 3$ [4,5], the change of electron density $\Delta n$ is related to a change of bias voltage $\Delta V$ by $\Delta n = \frac{\varepsilon_0 \varepsilon_{hBN}}{e*t} \Delta V$. Alternatively, we can use the oscillations in the reflectivity spectrum of the bright exciton as a function of gate voltage in the p-doped regime observed at +9 T (see Figure S1(b)). As demonstrated in Ref [6], these oscillations are due to the interaction of the exciton with the quantized Landau levels of the hole Fermi sea (see the sketch of Figure S1(a)). The period of the oscillations $\Delta V_{LL}$ is related to the filling of one Landau level $P_{LL} = \frac{eB}{2\pi\hbar}$ = 2.18 10$^{11}$ cm$^{-2}$. We can thus calculate the hole density as a function of the gate voltage by: $\Delta p = \Delta V \frac{P_{LL}}{\Delta V_{LL}}$. This yields the same estimation of the carrier density as the one deduced from the capacitance model. The advantage of this method is that it does not require knowledge of material parameters.

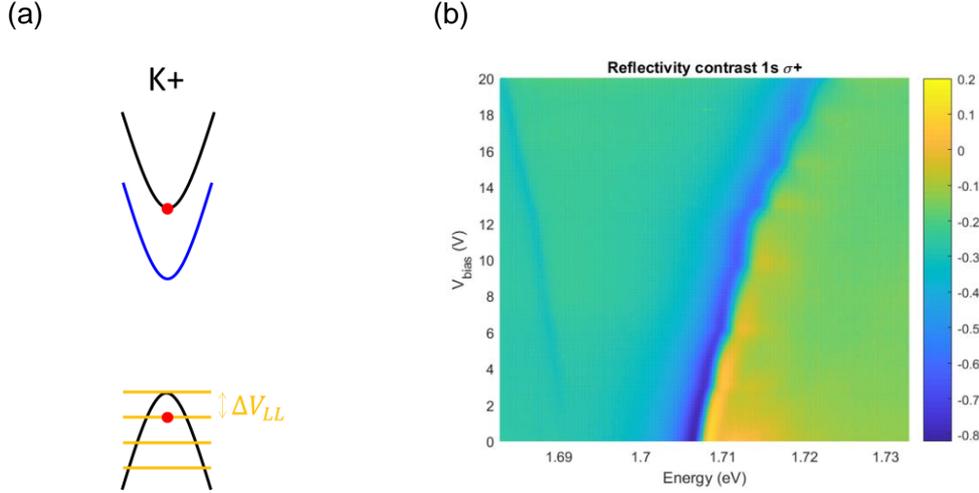

*Figure S1: Sketch of the band structure of the K+ valley at positive magnetic field showing the Landau levels. (b) Reflectivity contrast of the bright exciton with σ+ detection as a function of the gate voltage at +9 T*

Polarization dependent photoluminescence experiments are performed in close cycle cryostats (T= 4 K for $WSe_2$ and T=20 K for $WS_2$) with diffracted limited laser spot and continuous wave (cw) excitation. For $WSe_2$ we use the 632.8 nm line of a HeNe laser while a cw dye laser at a wavelength of 570 nm is used for $WS_2$. Polarization measurements are performed using a combination of Glan-Laser polarizers, quarter wave plate and half wave plate. The signal is dispersed by a monochromator and detected by a CCD camera.

The time-resolved photoluminescence (TRPL) measurements presented in the supplemental materials are performed in similar conditions: we used a ps-pulsed laser (TiSa) at a wavelength of 695 nm for $WSe_2$, and an OPO at 570 nm for $WS_2$. The signal is detected by a Hamamatsu streak camera with a time resolution of ~2-3 ps.

## S2. Estimation of the density of photo-generated electron-hole pairs

We can estimate the density of photo-generated bright electron-hole pairs (or excitons) from the excitation power density, the absorption coefficient and the lifetime of photo-generated electron-hole pairs. We have:

$$N_X = \frac{P_S \times \tau \times \alpha}{E_{photon}}$$

where $P_S$ is the excitation power density (500 W.cm$^{-2}$ for a power of 5 µW and a spot size of 1 µm$^2$), $E_{photon}$=1.96 eV for an excitation wavelength of 633 nm. A significant uncertainty comes from the absorption coefficient. For this non resonant wavelength, $\alpha$ was measured for free standing $WSe_2$ ML in the range 2-3% [4]. Assuming a lifetime $\tau$ of ~1 ps, this gives $N_X$~4x10$^7$ cm$^{-2}$, thus much smaller than the doping density.

## S3. Lifetime and spin relaxation time of trions

The degree of circular polarization of a state in a cw experiment is usually given by:

$$P = \frac{P_G}{1 + \frac{\tau}{\tau_s}}$$

where $P_G$ is the degree of circular polarization at the generation, $\tau$ is the lifetime of the state and $\tau_s$ is the spin (or valley) relaxation time (assuming here a single relaxation mechanism). In the simple model presented in the main text, we considered that the measured degrees of circular polarization for the triplet and the singlet directly reflect $P_G$. In other words we disregard the role of $\tau$ and $\tau_s$. We give some experimental justifications below. We performed time-resolved photoluminescence in a second $WSe_2$ charge tunable device and in the same $WS_2$ monolayer that is presented in the main text. In $WSe_2$ we measured the lifetime of both triplet and singlet as function of the doping density and in $WS_2$ we measured the decay of the circular polarization. Results are presented in Figure S2. At very low doping regime ($\sim 1 \times 10^{11}$ cm$^{-2}$), the triplet and the singlet luminescence decay with a lifetime $\tau$ of the order of 10-20 ps. When the doping increases to a few $10^{11}$ cm$^{-2}$, the decay is strongly reduced to $\sim$1 ps. We can have an idea of the spin relaxation time of trions $\tau_s$ by looking at the decay of the circular polarization of the triplet in $WS_2$ (Figure S2c). Our results suggest that this time is very slow (longer than 200 ps), thus much longer than the lifetime. In conclusion, it is reasonable to assume that $P \approx P_G$.

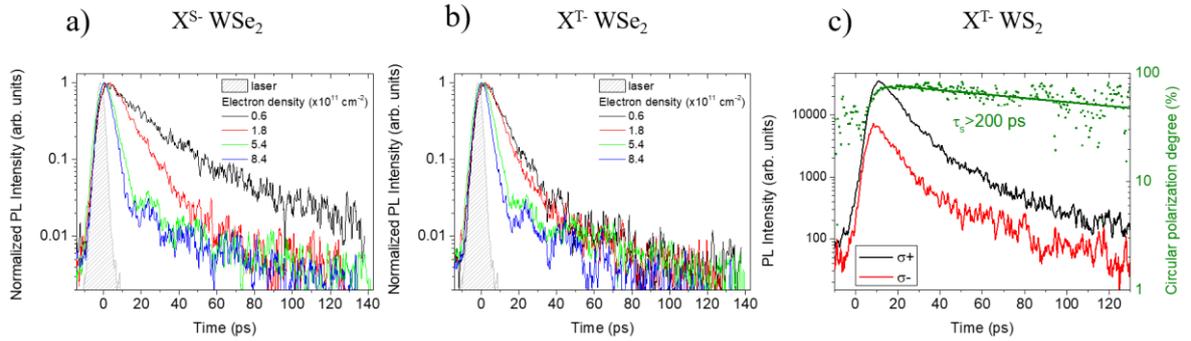

*Figure S2: Time-resolved photoluminescence measurements of the (a) singlet and (b) triplet in a second $WSe_2$ charge tunable device showing the shortening of the lifetime when the electron doping increases. (c) Time-resolved photoluminescence measurements of the triplet in the $WS_2$ monolayer sample of the main text. In this measurement, we excite with σ+ polarized light and detect both σ+ and σ- decays. In green we plot the circular polarization degree as a function of time. It shows that the spin/valley relaxation time of trions is much longer than the lifetime.*

## S4. Formation of bright trions through the binding of bright excitons and resident electrons: simple bimolecular model

In a simple bimolecular formation of trions we can write that the population of a given trion configuration $N_{X^{S,T-}}^{\sigma+,-}$ is proportional to the population of the photogenerated bright exciton $N_0^{K,K'}$ times the population of resident electrons $n_e^{K,K'}$. Thus we can write the degree of circular polarization for $X^{T-}$ and $X^{S-}$ as:

$$P_c(X^{T-}) = \frac{N_{X^{T-}}^{\sigma^+} - N_{X^{T-}}^{\sigma^-}}{N_{X^{T-}}^{\sigma^+} + N_{X^{T-}}^{\sigma^-}} \propto \frac{N_0^K n_e^{K'} - N_0^{K'} n_e^K}{N_0^K n_e^{K'} + N_0^{K'} n_e^K} \quad (S1)$$

$$P_c(X^{S-}) = \frac{N_{X^{S-}}^{\sigma^+} - N_{X^{S-}}^{\sigma^-}}{N_{X^{S-}}^{\sigma^+} + N_{X^{S-}}^{\sigma^-}} \propto \frac{N_0^K n_e^K - N_0^{K'} n_e^{K'}}{N_0^K n_e^K + N_0^{K'} n_e^{K'}} \quad (S2)$$

we can also define the degree of polarization of the photo-generated bright exciton (hot photo-generated excitons; or unbound electron-hole pairs; that eventually bind to resident electrons to form trions) as:

$$P_0 = \frac{N_0^K - N_0^{K'}}{N_0^K + N_0^{K'}} \tag{S3}$$

and the degree of polarization of resident electrons as:

$$P_e = \frac{n_e^K - n_e^{K'}}{n_e^K + n_e^{K'}} \tag{S4}$$

Combining the four equations we can show that:

$$P_c(X^{T-}) = \frac{P_0 - P_e}{1 - P_0 P_e} \tag{S5}$$

$$P_c(X^{S-}) = \frac{P_0 + P_e}{1 + P_0 P_e} \tag{S6}$$

Using the same model we can write the total PL intensities of triplet and singlet following circularly polarized excitation as:

$$I_{circ}(X^{T-}) \propto N_{X^{T-}}^{\sigma^+} + N_{X^{T-}}^{\sigma^-} \propto N_0^K n_e^{K'} + N_0^{K'} n_e^K \tag{S7}$$

$$I_{circ}(X^{S-}) \propto N_{X^{S-}}^{\sigma^+} + N_{X^{S-}}^{\sigma^-} \propto N_0^K n_e^K + N_0^{K'} n_e^{K'} \tag{S8}$$

For linear excitation, the total intensity is:

$$I_{lin}(X^{T-}) = I_{lin}(X^{S-}) \propto \frac{1}{2}(N_0^K + N_0^{K'})(n_e^K + n_e^{K'}) \tag{S9}$$

Thus the ratio of intensities between circular and linear excitation simply write:

$$R(X^{T-}) = \frac{I_{circ}(X^{T-})}{I_{lin}(X^{T-})} = 1 - P_0 P_e \tag{S10}$$

$$R(X^{S-}) = \frac{I_{circ}(X^{S-})}{I_{lin}(X^{S-})} = 1 + P_0 P_e \tag{S11}$$

In Figure S3, we plot $P_c(X^{T-})$, $P_c(X^{S-})$, $R(X^{T-})$ and $R(X^{S-})$ as a function of $P_0$ and $P_e$.

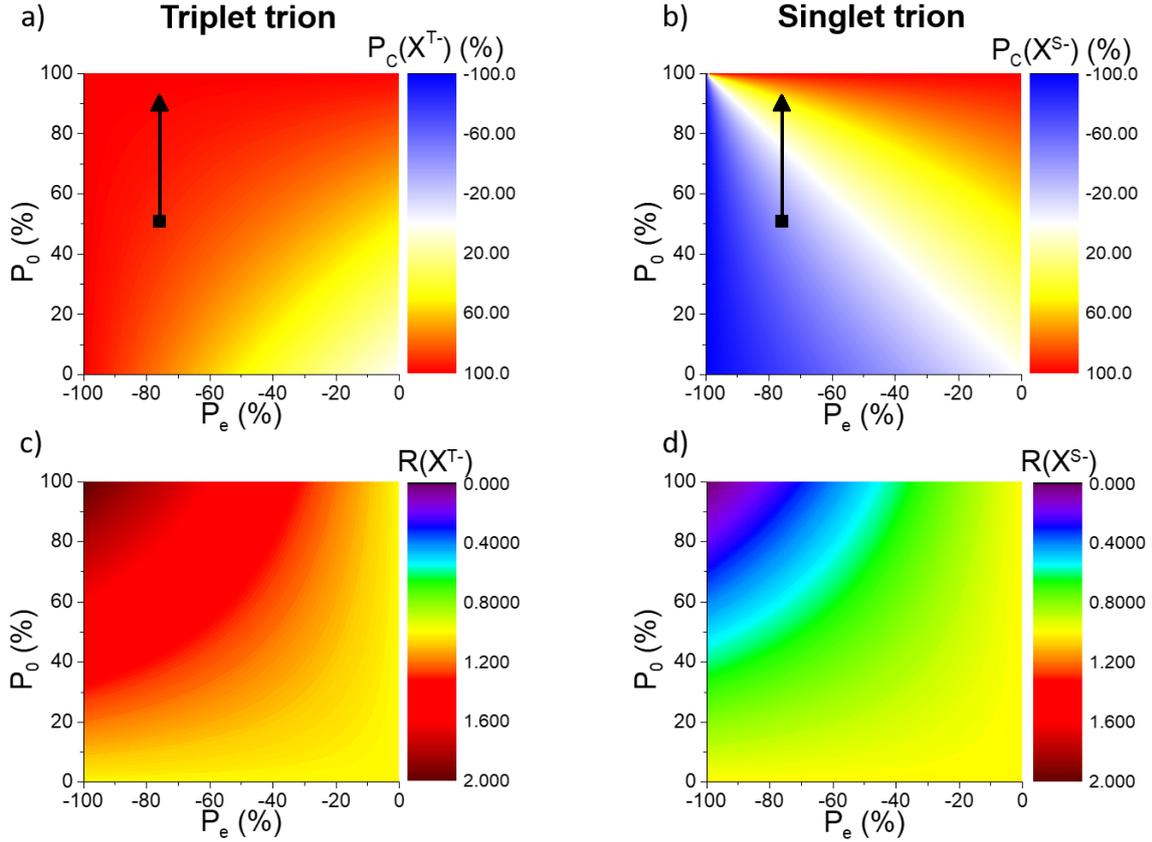

*Figure S3: **Bimolecular model**: Calculated degrees of PL circular polarization $P_c$ for (a) triplet and (b) singlet trions as function of the polarization of resident electrons ($P_e$) and the polarization of photogenerated excitons ($P_0$). (c) and (d) show the calculated ratio R for both triplet and singlet using the same model. The black data point correspond to the regime of Figure 2a of the main text where $P_c(X^{T-})=91\%$ and $P_c(X^{S-})=-40\%$. The vertical arrow correspond to the possible increase of $P_0$ when the doping decreases that may explain the results of Figure 2c where $P_c(X^{S-})$ turns positive and $P_c(X^{T-})$ remains very large.*

## S5. Additional data with increased excitation power

We show here the PL circular polarization degree and the ratio $R$ of both triplet and singlet trions as a function of the doping density for an increased excitation power of 20 µW. As compared to the measurements of Figure 2c and 2d (taken at 5 µW), the minimum of $P_c(X^{S-})$ is shifted above $4\times10^{11}$ cm$^{-2}$, and the maximum of $R(X^{T-})$ is found around $5\times10^{11}$ cm$^{-2}$ as compared to $3$-$4\times10^{11}$ cm$^{-2}$ at 5 µW. The overall shifts toward higher electron density is consistent with our scenario of dynamical polarization of resident electrons.

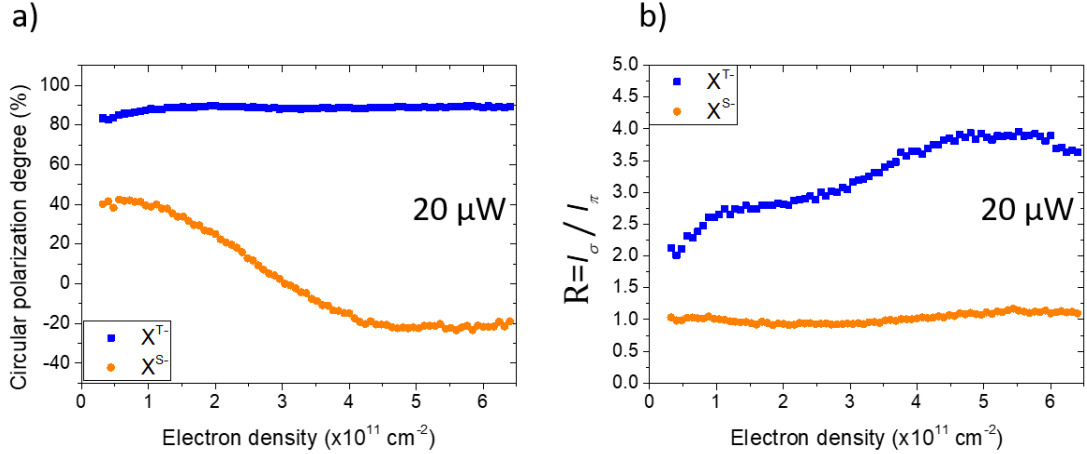

*Figure S4: (a) Circular polarization degree at the peak of triplet and singlet negative trions as a function of electron density for an excitation power of 20 µW. (b) Ratio of PL intensities between circular and linear excitations at the peak for both triplet and singlet as a function of electron density for an excitation power of 20 µW.*

## S6. Possible interpretation of the positive circular polarization of the singlet trion at low doping

In Figure 2c of the main text, we show that when the doping density decreases from $4\times10^{11}$ cm$^{-2}$, the negative circular polarization of the singlet trion $X^{S-}$ drops and even turns positive below $2\times10^{11}$ cm$^{-2}$. According to our scenario of dynamical building of polarization of resident electrons through cw circular excitation, there is no reason for the resident electrons to be less polarized at smaller doping densities. On the opposite, we showed that the polarization of resident electrons is favored when the density of photogenerated electrons is sufficient as compared to the doping density (see the power dependence of Figure 2e in the main text). Thus if we assume that the polarization of resident electrons $P_e$ remains at least $-76\%$ at low doping (the value we determined at $4\times10^{11}$ cm$^{-2}$ in the main text), our simple bimolecular model dictates that the polarization of $X^{S-}$ can turn positive and the polarization of $X^{T-}$ remains very large if $P_0 > -P_e$ (i.e. $P_0 > 76\%$) (see equation (2) of the main text and Figure S3). In other words, our results can be explained if the polarization of photo-generated excitons $P_0$ increases when the doping density decreases.

Note that $P_0$ does not directly correspond to the polarization of bright excitons $X^0$ measured in PL ($P_c(X^0) \sim 15\%$ in Figure 2a). Indeed, $P_c(X^0)$ corresponds to cold bright excitons (i.e. radiatively recombining in the light cone) while $P_0$ corresponds to hot photo-generated excitons (or even unbound electron-hole pairs) that bind to resident electrons to form trions. In a simple approach we can write:

$$P_c(X^0) = \frac{P_0}{1 + \dfrac{\tau^{X^0}}{\tau_s^{X^0}}}$$

where $\tau^{X^0}$ and $\tau_s^{X^0}$ are the bright exciton lifetime and spin relaxation time. When the doping decreases, we can assume that $\tau^{X^0}$ remains constant or increases and that $\tau_s^{X^0}$ remains constant or decreases (the long range exchange interaction may be less efficient at higher doping because of the screening by the Fermi sea). In Figure S5, we show that $P_c(X^0)$ increases up to 35% when the electron doping decreases. Thus it is consistent with an increase of $P_0$ when the doping decreases.

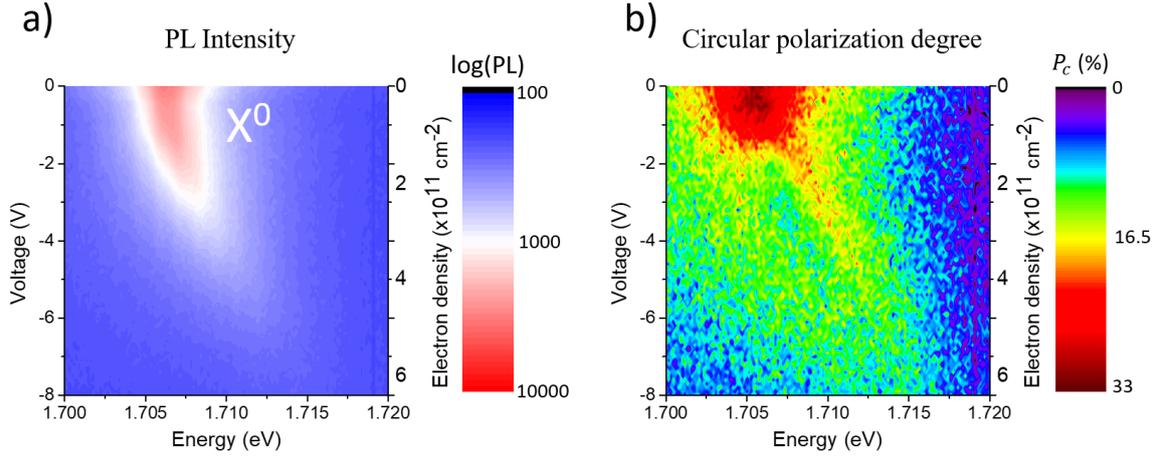

*Figure S5: (a) Zoom on the PL intensity of the bright exciton $X^0$ as a function of electron density. (b) Corresponding circular polarization degree showing an increase of $P_c(X^0)$ when the doping decreases.*

## S7. Alternative formation mechanisms of trions

In the main text we considered that bright trions are formed through a simple bimolecular mechanism involving a bright exciton and bottommost conduction band electrons. We briefly detail in this section alternative formation mechanisms.

- triplet trions can be formed through the binding of indirect excitons and topmost conduction band electrons.

- singlet trions can be formed through the binding of dark excitons and topmost conduction band electrons.

- triplet trions can convert to singlet trions through a second-order exchange process. The triplet has higher energy than the singlet by $\Delta_{TS}$ ~6 meV, and therefore, an electron-hole exchange of the bright exciton component in the trion cannot alone switch between triplet and singlet trion states. However, the conversion from the higher energy trion (triplet) to the lower-energy one (singlet) can proceed if the electron-hole exchange is accompanied by Coulomb scattering of the trion with a cold resident electron. Here, the excess of $\Delta_{TS}$ ~6 meV are mostly gained by the resident electron after scattering. The singlet-to-triplet conversion is suppressed because this would require the resident electron to give away 6 meV, which is not possible if the electron is cold (resides in the bottom of the valley).

- triplet trions can become dark trions following emission of a zone-center phonon $\Gamma_5$.

- singlet trions can become dark trions following emission of a zone-edge phonon $K_2$.

- bright trions can turn to dark through trion-electron Coulomb scattering. Here, an incoming free electron kicks the top-valley electron from the incoming bright trion complex, and binds to the left-behind exciton component. The end products are a top-valley electron and a bound dark trion with two electrons in time-reversed bottom valleys. The energy conservation of this process mandates that the top-valley free electron after the scattering gains kinetic energy compared to the kinetic energy of the incoming electron. Importantly, this bright-to-dark trion conversion is suppressed when the free resident electrons are strongly polarized (e.g., resident electrons reside mostly in K' and the bottom-valley electron of the triplet or singlet is also in K'). This mechanism is thus consistent with the positive polarization of the triplet and the negative polarization of the singlet.

- When the electrostatic doping is relatively large, a singlet (triplet) trion may be formed through a trimolecular binding process wherein a photoexcited hole, photoexcited electron (top-valley), and a resident electron (bottom valley) bind to form a bright trion. This process requires a relatively large density of resident electrons in the bottom valleys, so that trions are formed before the top-valley electrons relax to the bottom valleys. This mechanism is detailed in the next section.

## S8. Trimolecular model

We show here that the results of Figure 2a and 2b of the main text can be interpreted using a model taken into account a trimolecular formation of bright (singlet and triplet) and dark negative trions. We detail the model below. We use the notations of Figure S6 to describe the populations of electrons, holes and trions in each valley.

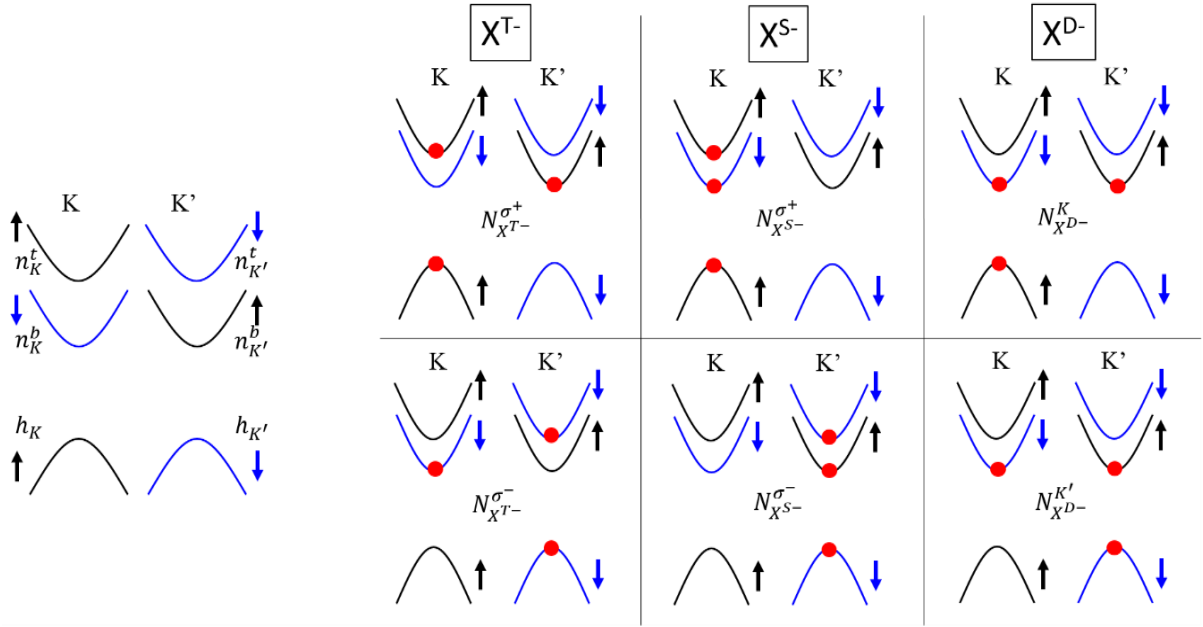

*Figure S6: Notations of the different electron, hole and trion populations used in the rate equation model. The arrows represent the electron conduction or valence spin states in each valley.*

For the sake of simplicity, we consider in our model only the formation and recombination mechanisms of trions (no spin relaxation). This can be easily justified by the very short lifetime of the trions as compared to the spin relaxation times (see Figure S2). We also consider that the dynamic polarization of resident electrons takes place at a much slower timescale than the formation and recombination of the trions so that $n_K^b$ and $n_{K'}^b$ are fixed (they are considered as parameters of the model). We have $n_K^b + n_{K'}^b = n_e$ where $n_e$ is the doping density and $P_e = \frac{n_K^b - n_{K'}^b}{n_K^b + n_{K'}^b}$ is the polarization of the resident electrons.

Within these assumptions, the different electron and hole populations satisfy the following system of rate equations:

$$\frac{dn_K^t}{dt} = G_K^0 - C_{triplet} n_K^t n_{K'}^b h_K - C_{singlet} n_K^t n_K^b h_K - \frac{n_K^t}{\tau_0}$$

$$\frac{dn_{K'}^t}{dt} = G_{K'}^0 - C_{triplet} n_{K'}^t n_K^b h_{K'} - C_{singlet} n_{K'}^t n_{K'}^b h_{K'} - \frac{n_{K'}^t}{\tau_0}$$

$$\frac{dh_K}{dt} = G_K^0 - C_{triplet} n_K^t n_{K'}^b h_K - C_{singlet} n_K^t n_K^b h_K - C_{dark} n_K^b n_{K'}^b h_K - \frac{h_K}{\tau_0}$$

$$\frac{dh_{K'}}{dt} = G^0_{K'} - C_{triplet}n^t_{K'}n^b_K h_{K'} - C_{singlet}n^t_{K'}n^b_{K'}h_{K'} - C_{dark}n^b_{K'}n^b_K h_{K'} - \frac{h_{K'}}{\tau_0}$$

and the different populations of trions read as:

$$\frac{dN^{\sigma^+}_{X^{T-}}}{dt} = C_{triplet}n^t_K n^b_{K'} h_K - \frac{N^{\sigma^+}_{X^{T-}}}{\tau_T}$$

$$\frac{dN^{\sigma^-}_{X^{T-}}}{dt} = C_{triplet}n^t_{K'} n^b_K h_{K'} - \frac{N^{\sigma^-}_{X^{T-}}}{\tau_T}$$

$$\frac{dN^{\sigma^+}_{X^{S-}}}{dt} = C_{singlet}n^t_K n^b_K h_K - \frac{N^{\sigma^+}_{X^{S-}}}{\tau_S}$$

$$\frac{dN^{\sigma^-}_{X^{S-}}}{dt} = C_{singlet}n^t_{K'} n^b_{K'} h_{K'} - \frac{N^{\sigma^-}_{X^{S-}}}{\tau_S}$$

$$\frac{dN^{K}_{X^{D-}}}{dt} = C_{dark}n^b_K n^b_{K'} h_K - \frac{N^{K}_{X^{D-}}}{\tau_D}$$

$$\frac{dN^{K'}_{X^{D-}}}{dt} = C_{dark}n^b_{K'} n^b_K h_{K'} - \frac{N^{K'}_{X^{D-}}}{\tau_D}$$

where $C_{triplet}$, $C_{singlet}$, and $C_{dark}$ are the trimolecular formation rates of $X^{T-}$, $X^{S-}$ and $X^{D-}$. $\tau_T$, $\tau_S$, $\tau_D$ and $\tau_0$ are the lifetimes of $X^{T-}$, $X^{S-}$, $X^{D-}$ and $X^0$. $G^0_K$, $G^0_{K'}$ are the generation rates of the photogenerated electron-hole pairs in each valley.

- We fix $\tau_0$=1 ps, $\tau_T$=$\tau_S$=10 ps and $\tau_D$=500 ps as measured in time-resolved photoluminescence experiments (see Figure S2 and Figure S7). Note that except for $\tau_0$, these parameters do not change the values of $P_c(X^{T-})$, $P_c(X^{S-})$, $R(X^{T-})$ and $R(X^{S-})$.
- $G^0_K$ and $G^0_{K'}$ are defined through the total generation rate of electron-hole pairs $G^0 = \frac{P_S \times \alpha}{E_{photon}}$ and $P_0 = \frac{G^K_0 - G^{K'}_0}{G^K_0 + G^{K'}_0}$ where $\alpha$ is the absorption coefficient and $P_S$ is the excitation power density.
- We fix $C_{triplet} = C_{singlet} = C_B$ as we expect the formation coefficient of triplet and singlet to be similar given that they have very similar oscillator strength (see reflectivity spectra in Figure 1b of the main text and [5]).

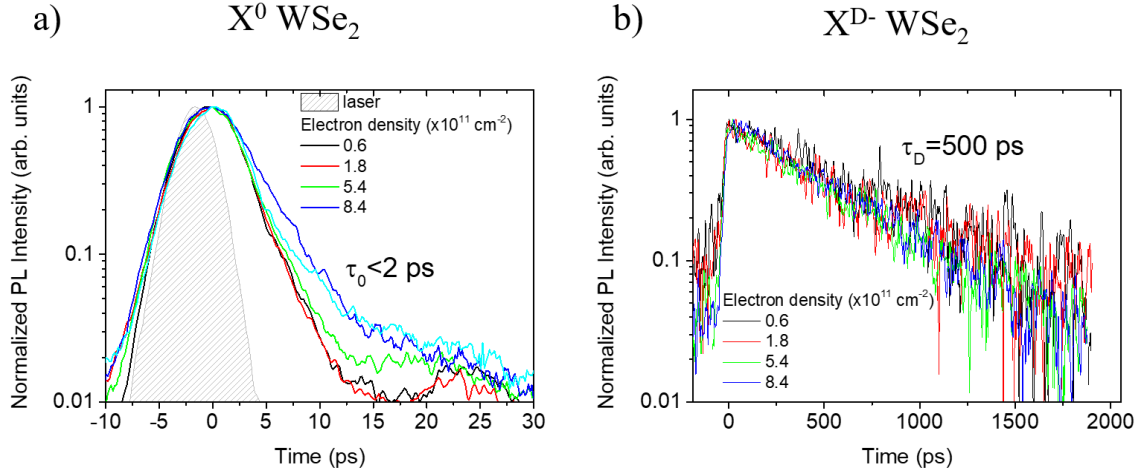

*Figure S7: Time-resolved photoluminescence measurements of the (a) bright exciton and (b) dark trion in a second WSe$_2$ charge tunable device. The bright exciton lifetime is close to the temporal resolution. We chose to fix $\tau_0$ in agreement with the values found in the literature [ref].*

The different populations for both circular and linear excitations are calculated by solving the equations system in stationary conditions ($\frac{d}{dt}=0$). We fit simultaneously $P_c(X^{T-})$, $P_c(X^{S-})$, $R(X^{T-})$ and $R(X^{S-})$ to the experimental values of Figure 2 of the main text ($P_c(X^{T-})$=91%, $P_c(X^{S-})$=−40%, $R(X^{T-})$=4.4 and $R(X^{S-})$=0.88) by adjusting 5 parameters: $P_0$, $P_e$, $\alpha$, $C_B$ and $C_D$

We obtain $P_c(X^{T-})$=98%, $P_c(X^{S-})$=−40%, $R(X^{T-})$=4.4 and $R(X^{S-})$=0.9 for $P_0$=57%, $P_e$=−88%, $\alpha$=2.95%, $C_B$=4.65x10$^{-17}$ cm$^{-4}$ps$^{-1}$, $C_D$=4.8x10$^{-20}$ cm$^{-4}$ps$^{-1}$.

$P_0$ and $P_e$ are in the same range than the values obtained with the simple bimolecular model ($P_0$=51% and $P_e$=−76%). We show in Figure S8, the influence of these two parameters on the results of the trimolecular model.

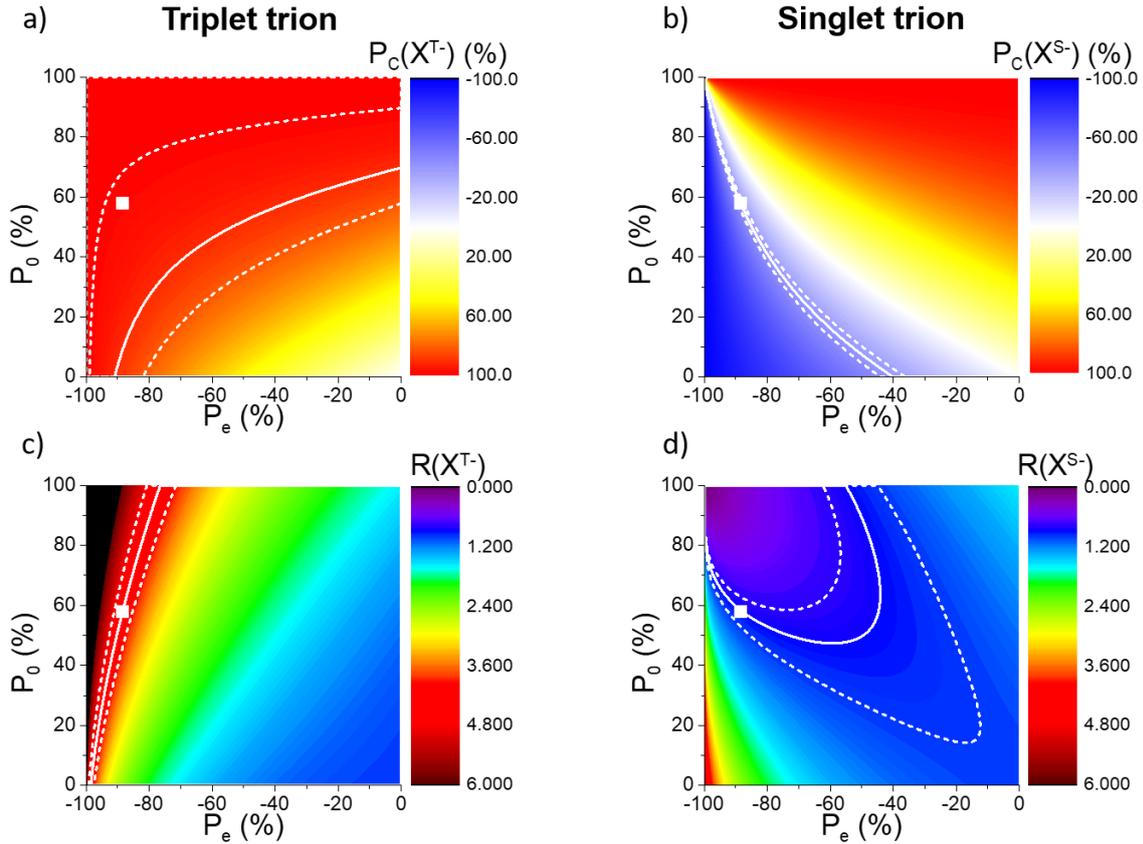

*Figure S8: **Trimolecular model**: Influence of $P_0$ and $P_e$ on the calculated degrees of circular polarization for (a) triplet and (b) singlet trions and on the ratio R (c) and (d). The white data point correspond to the result of the fit ($P_0=57\%$, $P_e=-88\%$). The white solid lines are the contour plot of the experimental data ($P_c(X^{T-})=91\%$, $P_c(X^{S-})=-40\%$, $R(X^{T-})=4.4$ and $R(X^{S-})=0.88$) and the dashed lines are the contour plots of the experimental data with a tolerance of ±10%. All other parameters are fixed to the values written in the text.*

The absorption coefficient $\alpha$ is also perfectly consistent with the literature [4] and the value of 3% that we used in section S2 to estimate the exciton density. We show in Figure S9, the influence of both $\alpha$ and $\tau_0$ on the results of the trimolecular model. It shows that $\alpha$ and $\tau_0$ are linked (a larger $\alpha$ requires a smaller $\tau_0$ to give the same results).

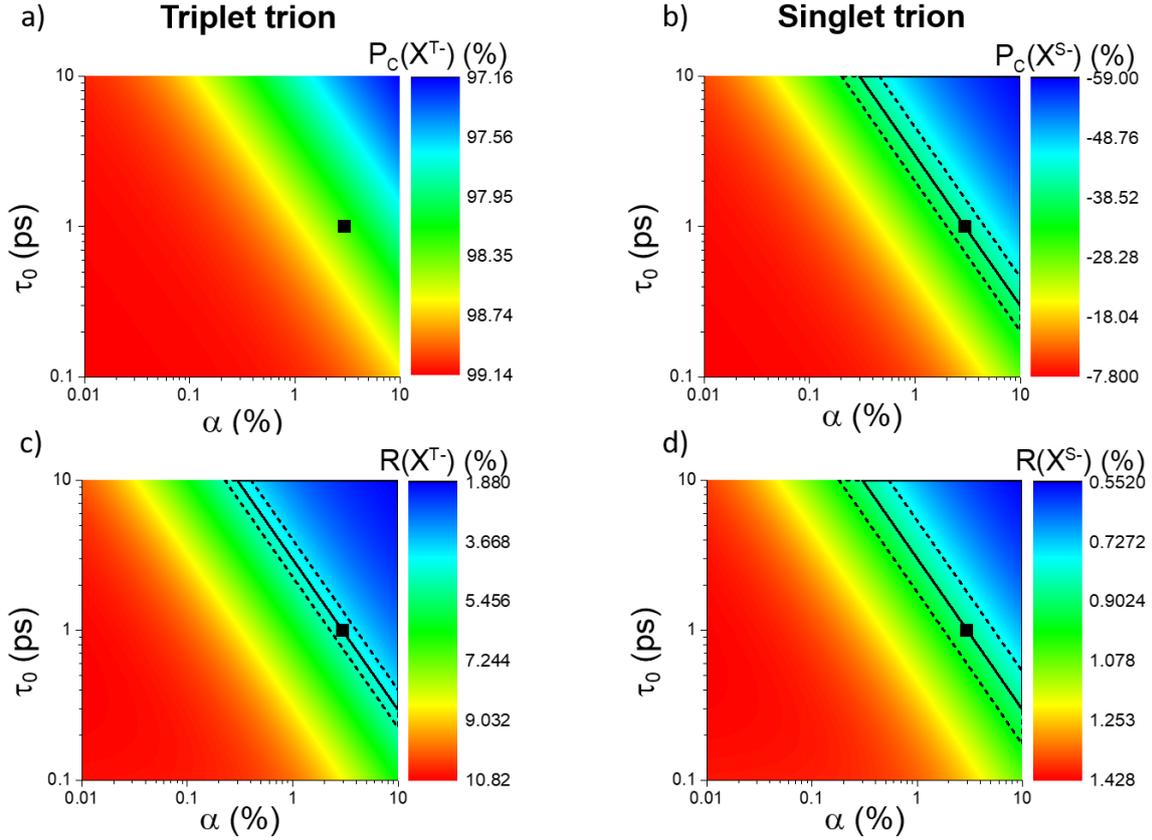

*Figure S9: **Trimolecular model**: Influence of the absorption coefficient α and the bright exciton lifetime $\tau_0$ on the calculated degrees of circular polarization for (a) triplet and (b) singlet trions and on the ratio R (c) and (d). The black data point correspond to the result of the fit. The black solid lines are the contour plot of the experimental data ($P_c(X^{T-})$=91%, $P_c(X^{S-})$=-40%, $R(X^{T-})$=4.4 and $R(X^{S-})$=0.88) and the dashed lines are the contour plots of the experimental data with a tolerance of ±10%. All other parameters are fixed to the values written in the text.*

Finally, we discuss about the trimolecular formation coefficients. We show in Figure S10, the influence of $C_B$ and $C_D$ on $P_c(X^{T-})$, $P_c(X^{S-})$, $R(X^{T-})$ and $R(X^{S-})$. We clearly see that for $C_D > 10^{-21}$ cm$^{-4}$.ps$^{-1}$, the results are only sensitive to the ratio $C_B/C_D$ which in our fit is close to 1000. Interestingly, this ratio also fix the ratio between the PL intensity of bright and dark trions. In Figure 2b of the main text we observed that $X^{T-}$, $X^{S-}$ and $X^{D-}$ have a similar intensity for linear excitation. With our model, we find that $X^{D-}$ is 10 times larger than $X^{T-}$ and $X^{S-}$ which is not so far from our experimental observations (given that the light emitted by $X^{D-}$ is not well collected due to its in-plane direction). We would also like to mention that with our model, the total PL intensity (bright trions + dark trions) is the same for circular and linear excitation.

In conclusion, despite its simplicity, our model describes rather well our experimental observations with only a few set of parameters with values perfectly consistent with their scientific meaning.

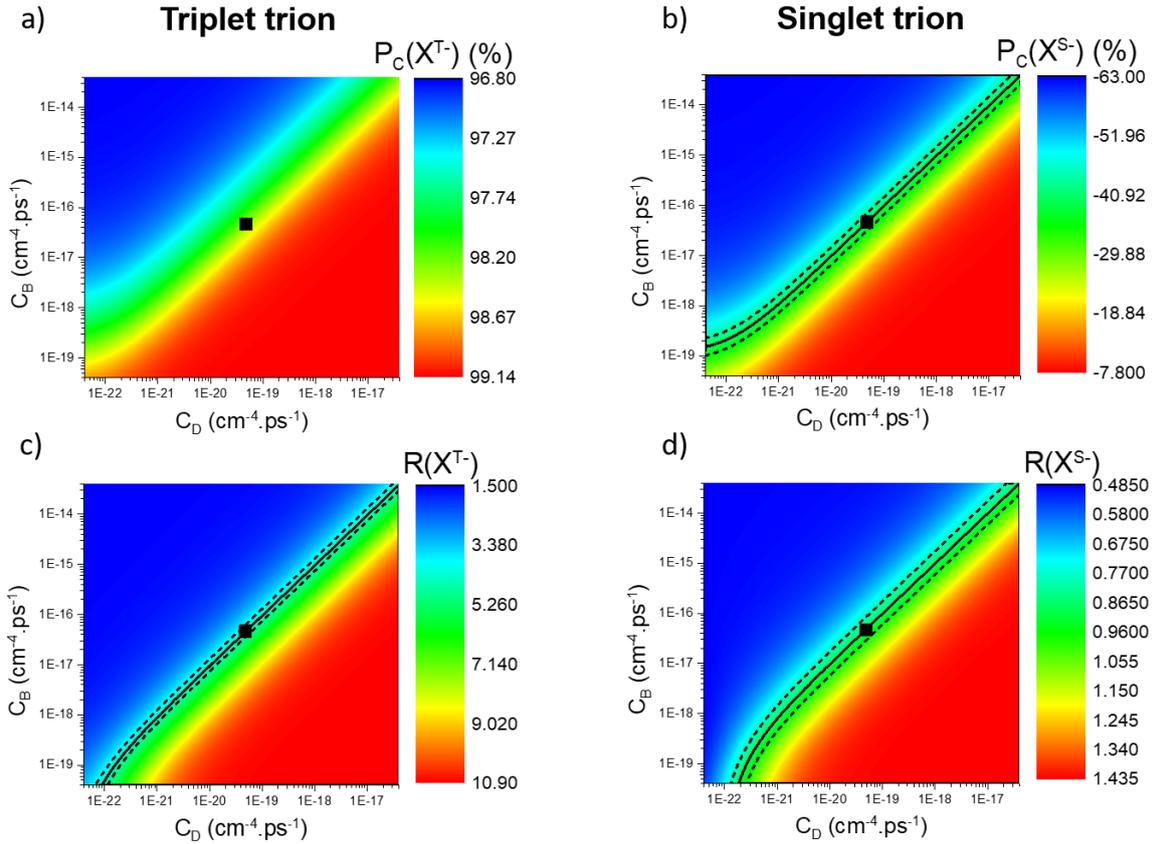

*Figure S10: **Trimolecular model**: Influence of the trimolecular formation coefficients of the bright and dark trions $C_B$ and $C_D$ on the calculated degrees of circular polarization for (a) triplet and (b) singlet trions and on the ratio R (c) and (d). The black data point correspond to the result of the fit. The black solid lines are the contour plot of the experimental data ($P_c(X^{T-})$=91%, $P_c(X^{S-})$=-40%, $R(X^{T-})$=4.4 and $R(X^{S-})$=0.88) and the dashed lines are the contour plots of the experimental data with a tolerance of ±10%. All other parameters are fixed to the values written in the text.*

## S9. Power dependence in WS$_2$

We present the circular polarization degree $P_c$ and the ratio $R$ for both triplet and singlet trions in WS$_2$ ML as function of laser power. The results are very similar to WSe2 (Figure 2e and f of the main text): when the excitation power decreases, $P_c(X^{T-})$ and $P_c(X^{S-})$ converge to the same value (around 30% for WS$_2$) and $R(X^{T-})$ and $R(X^{S-})$ converge to 1.

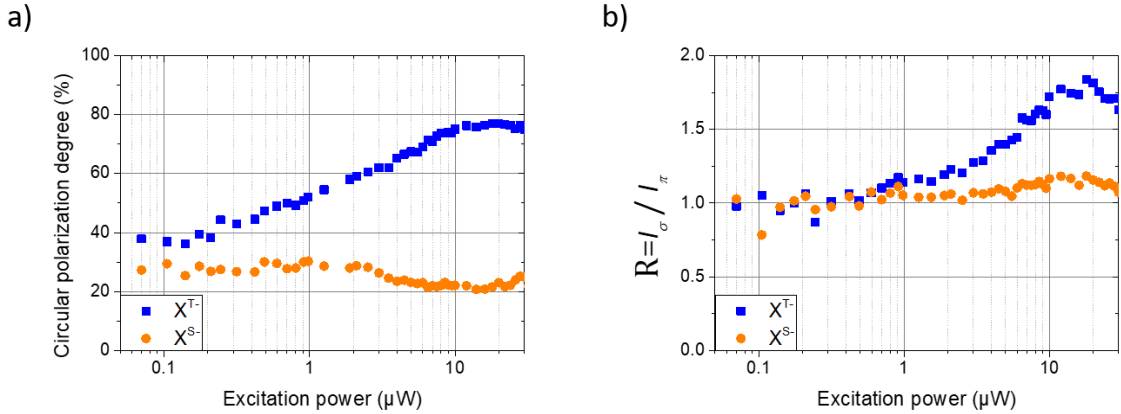

*Figure S11: (a) Circular polarization degree at the peak of triplet and singlet as a function of excitation power. (b) Ratio of PL intensities between circular and linear excitations at the peak for both triplet and singlet as a function of excitation power.*